\documentclass[a4paper,11pt]{article}
\usepackage{graphicx}
\usepackage{graphicx}
\usepackage{psfrag}
\usepackage{subfigure}
\usepackage{amsmath,amstext,amsfonts,amsbsy,amssymb,amscd,bbm,epsfig}
\usepackage{verbatim}
=-0.5cm \oddsidemargin=-0.5cm
\usepackage[utf8x]{inputenc}

\title{Exploration of the phase diagram of 5D anisotropic SU(2) gauge theory}
\date{}
\author{K.~Farakos\footnote{E-mail: kfarakos@central.ntua.gr} ~and
S.~Vrentzos\footnote{E-mail: vrentsps@central.ntua.gr}}
\begin{document}
\maketitle
\begin{center}
Physics Department, National Technical University
of Athens,
\\ Zografou Campus 15780, Greece
\end{center}

\begin{abstract}
In this paper we attempt a non-perturbative study of the five dimensional, anisotropic SU(2) gauge theory on the lattice using Monte-Carlo techniques.
Our goal is the exploration of the phase diagram, define the various phases and the critical boundary lines. Three phases appear, two of them
are continuations of the Strong and the Weak coupling phases of pure 4d SU(2) to non-zero coupling $\beta^{'}$ in the fifth
transverse direction and they are separated by a crossover transition, while the third phase is a 5d Coulombic phase. We provide evidence
that the phase transition between the 5D Coulomb phase and the
Weak coupling phase is a second order phase transition. Assuming that this result is not
altered when increasing the lattice volume we give a first estimate of the
associated critical exponents.
This opens the possibility
for a continuum effective five dimensional field theory.
\end{abstract}

\newpage
\section{Introduction}
Models with extra dimensions have been used many times in the physics of elementary particles and cosmology.
The first attempt has been in 1922 by Kaluza and Klein in an attempt to unify all the known interactions at this time, i.e.
electromagnetism and gravity \cite{KK}. String theories and M-theory, that also incorporate quantum gravity effects, consider space times
with extra spatial dimensions to be consistently defined. In these theories ten (String theory) or eleven (M-theory)
space time dimensions are required.
 One of the main problems with the extra dimensional models is their connection with the four dimensional physics.
 On the other hand gauge theories are non-renormalizable in higher dimensions and we have to define them with a cutoff $\Lambda$. We can interpret
 these models as a low-energy effective description of an underlying more fundamental theory. The effective theory is then valid
 for energies lower than the cutoff $\Lambda$. After this energy scale the underlying theory becomes relevant.
 The lattice regularization gives a gauge invariant cutoff for these models and provides a natural setup for a non-perturbative study.\\
 Today the main approach to the models with higher dimensions is through the "brane" models. We suppose that our 4d world (brane) is embedded in a
 higher dimensional space with n extra dimensions, D=4+n. It is possible to consider flat extra dimensions (ADD model) \cite{Anton} or curved ones
 like the Randall-Sundrum model \cite{RS}. Gravity is supposed to propagate freely in the higher dimensional manifold (bulk) and the Standard Model
 particles are pinned on the brane. In this respect we need a mechanism for localization
 of them on the 4d brane see for example \cite{FRPP1} and references there in.\\
 On the lattice such a mechanism has been proposed by Fu and Nielsen \cite{FN1} in 1984. They suggested that if in a D dimensional lattice
 we define a set of different nearest-neighbor couplings for the gauge fields in a d dimensional sublattice, this may lead to the formation
 of a d-dimensional Layer phase ($d<D$). Within a d-dimensional layer phase, gauge particles and charged particles can travel as if they were
 massless. In any attempt to probe the remaining n dimensions, $n=D-d$, the above particles propagate as if they were confined. Actually,
 it is this confinement which forbids the interaction with neighboring d dimensional layers and implies the effective detection of a
 four-dimensional world.\\
 In a pure anisotropic D=5 U(1) gauge theory, with different couplings $\beta$ and $\beta^{'}$ in a 4d subspace and the extra
 fifth transverse direction
 correspondingly we observe three phases: the Strong-Confining phase, the Layer phase with Coulomb interaction in the four dimensional layers and a
 five dimensional Coulomb deconfined phase \cite{FV2,KA,DF}. Note that the transition line between the Strong coupling ($\beta<1$) phase and the Layer phase
 ($\beta>1, \beta^{'}<0.4$) is weakly first order. The phase transition between the Layer phase and the five-dimensional Coulomb phase,
 $\beta^{'}>0.4$, is probably of second order \cite{DFV}, while the transition from the Strong coupling to 5D Coulomb phase is a strong first order
 phase transition.
 For the Abelian Higgs model and the SU(2) Adjoint Higgs model in five dimensions lattice Monte Carlo results give serious evidence for a layered
 structure \cite{DF2}. Finally we would like to mention that the layer phase disappears at high temperature \cite{FVFT}. Note that the layer structure in the previous models
 is based on the fact that when $\beta^{'}$ in the extra dimension is zero then there is a non-trivial phase transition.\\
 In this paper we study the phase diagram for the anisotropic five dimensional pure SU(2) gauge theory. The interest in studying five dimensional pure gauge theories is related to a particular extension of the Standard Model which is
 called gauge-Higgs unification. The idea behind this model is the identification of the Higgs boson with the extra dimensional components
 of the gauge field as the system undergoes a dimensional reduction from five to four dimensions \cite{Irges,FdF,KnechtliRago}.
 A first step in this study for the dimensional reduction is to explore the phase diagram of the theory.
 Four dimensional pure SU(2) gauge theories have a unique phase in which interactions are confined. In the phase diagram there is only one fixed point for
 $\beta\rightarrow\infty$ ($g^{2}\rightarrow0$); this is an ultraviolet fixed point where we can use weak coupling perturbation theory. So the
 confining strong coupling region and the perturbative weak coupling region describe the same physics. There is a crossover transition
 in the lattice phase diagram in four dimensions ($\beta^{'}=0$) for $\beta\approx2.20$. We expect that this behavior continues also for
 $\beta^{'}>0$ and there is no Layer phase in this model.\\
 The main scope of this work is to determine by Monte Carlo methods the order of the phase transition to the five dimensional
 Coulomb deconfining phase. Mean field results show that this phase transition is of first order for large $\beta^{'}$ and small $\beta$ values,
 and becomes a second order one for small $\beta^{'}$ and large $\beta$  values\cite{Irges}. It has been also proposed that the
 second order phase transition belongs to the 4d Ising universality class \cite{Irges,FdF}.\\
 In section 2 we write down the lattice action for the model and we define the order parameters that we use for the analysis of the phase diagram.
 In section 3 we present the phase diagram and the order of the possible phase transitions. We also give estimates for the values of the critical exponents for the observed bulk continuous transitions.

\section{The model}
We study the five dimensional SU(2) Yang-Mills theory in  5D Euclidean spacetime. The model is formally defined through the relations:
\begin{equation}
 Z=\int{DA} e^{-S_{E}} \quad \textnormal{with}\quad S_{E}=\int{d^{4}x}\int{dx_{5}}\frac{1}{2g^{2}_{5}}TrF^{2}_{MN}
\end{equation}
 were the upper case indices M,N refer to the 5D space and $g_{5}$ is the dimensionful gauge coupling in five dimensions.\\
The regularization of the model on a five dimensional anisotropic lattice with lattice spacing $a_{s}$ in four dimensions and $a_{5}$
for the fifth one, is given in terms of the action
\begin{equation}
 S_{E}^{L}=\beta\sum_{x}\sum_{1\leq\mu<\nu\leq 4}{(1-\frac{1}{2}TrU_{\mu\nu}(x))}+\beta^{'}\sum_{x}\sum_{1\leq\mu\leq 4}{(1-\frac{1}{2}TrU_{\mu 5}(x))}
\end{equation}
with $U_{\mu}=e^{i a_{s}A_{\mu}}$, $U_{5}=e^{i a_{5}A_{5}}$ and $\beta$, $\beta^{'}$ denote the couplings in the 4d subspace and along the extra (transverse) dimension respectively.
The gauge field is represented by 2x2 Hermitian matrices $A_{\mu}=A_{\mu}^{\alpha}\sigma^{\alpha}$ where $\sigma^{\alpha}$ are the Pauli matrices.
$U_{\mu\nu}$ and $U_{\mu5}$ are the plaquettes defined on the four dimensional space and along
the extra fifth (transverse) direction $x_{5}$. Note that $\beta=\frac{4 a_{5}}{g^{2}_{5}}$,
$\beta^{'}=\frac{4 a^{2}_{S}}{g^{2}_{5} a_{5}}$ in the classical limit.\\

The order parameters used in this study are either space-like or transverse-like ones with the following definitions:

\begin{equation}
 \textnormal{Space-Plaquette:}\quad \hat{P}_{S}\equiv \frac{1}{6L^{5}}\sum_{x,1\leq\mu<\nu\leq4}TrU_{\mu\nu}(x)
\end{equation}
\begin{equation}
 \textnormal{Transverse-Plaquette:}\quad \hat{P}_{5}\equiv \frac{1}{4L^{5}}\sum_{x,1\leq\mu\leq4}TrU_{\mu5}(x)
\end{equation}
with L the linear dimension of the lattice $(V_{5D}=L^{5})$.\\
Starting from the operators in Eqs.~(3) and (4) we measure the space-like plaquette mean value \begin{math}P_{S} \equiv \left < \hat{P}_{S}\right>\end{math} and the
transverse-like plaquette mean value \begin{math}P_{5}\equiv \left< \hat{P}_{5}\right>\end{math}.
The symbol $\left <\dots \right >$ denotes the statistical average with respect to the action given by Eq.~(2).\\
We also make use of the corresponding susceptibilities:
\begin{equation}
 S(\hat{P}_{S})=V_{5D}\Big{(}\left<\hat{P}^{2}_{S}\right>-\left<\hat{P}_{S}\right>^{2}\Big{)}\quad \textnormal{ and }\quad S(\hat{P}_{5})=V_{5D}\Big{(}\left<\hat{P}^{2}_{5}\right>-\left<\hat{P}_{5}\right>^{2}\Big{)}
\end{equation}
and the distributions $N(\hat{P}_{S}), N(\hat{P}_{5})$ of $\hat{P}_{S}$ and $\hat{P}_{5}$ respectively.

\section{Lattice Results}

\subsection{Monte-Carlo details}
In our simulations we used the Kennedy-Pendleton heat bath algorithm for the updating of the gauge field \cite{KP}. A sketch of the phase diagram was produced from
simulations on a lattice volume $V=8^5$. However the identification of the various phases was done using much larger volumes.
The critical points in the ($\beta^{'}$,$\beta$) plane correspond to the peak of the
relevant (space-like or transverse-like) susceptibility.
For the study of the order of the transitions between the phases of the model we used volumes up to $16^5$
with high statistics runs. For the errors the jack-knife method has been used.

In the following we present first the phase diagram of the theory with a short description of the various phases and  continue with a more detailed study of the model for different
values of the gauge couplings $\beta$ and $\beta^{'}$.
\subsection{The phase diagram}

\begin{figure}[ht!]
\begin{center}
\includegraphics[scale=0.58]{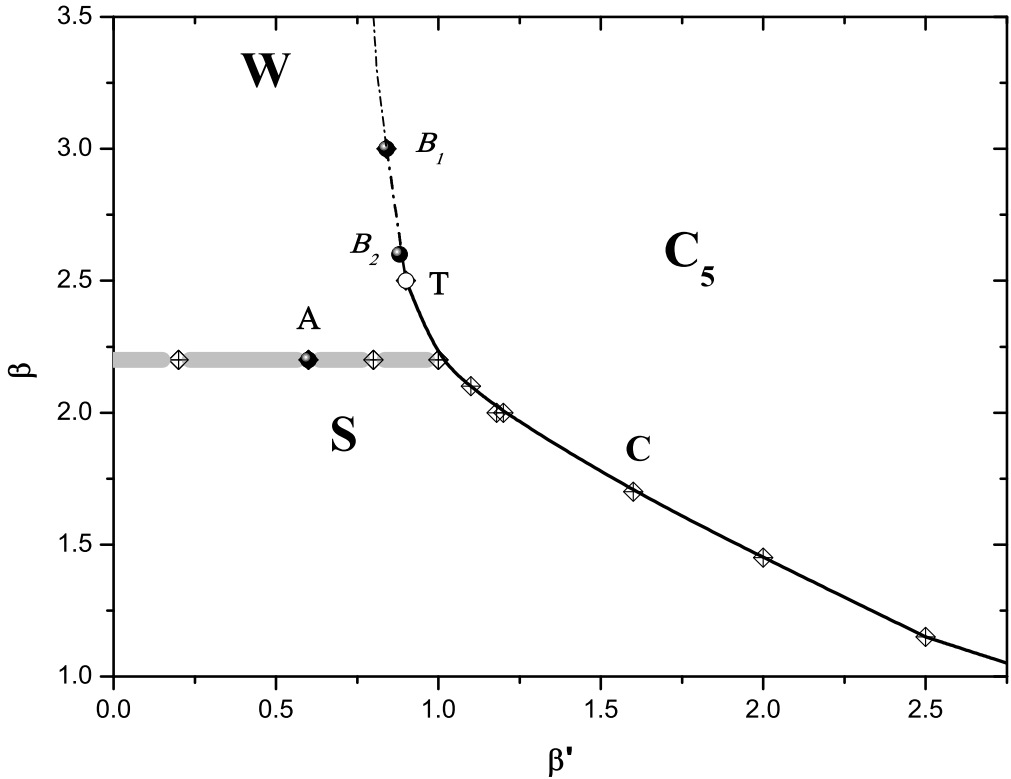}
\end{center}
\caption{The phase diagram for the 5D anisotropic SU(2) gauge model. $\bold{{C}_{5}}$ is the 5D Coulomb phase, $\bold{S}$ refer to the strongly coupled confining phase and the third phase $\bold{W}$ is the continuation of the weak coupling phase of 4d SU(2) for non zero $\beta^{'}$. The Coulomb phase $\bold{C_{5}}$ is completely separated from the other two phases $\bold{S}$ and $\bold{W}$. For details about the various phases and the order of the phase transitions see the text.}
\label{3-1-1}
\end{figure}
The phase diagram that we present in Fig.~1 has been produced using a number of measurements ranging
 from $10^{4}$ to $10^{5}$ (after thermalization) on $V=8^{5}$ volumes. The phase diagram consists of three phases.
For pure gauge theories analytical predictions regarding the value of
the plaquette in the strong and weak
coupling regime are useful for the characterization of the various phases.
We know that for the isotropic SU(2) model defined in D dimensions the plaquette in the strong
coupling limit ($\beta_{g} \ll 1$) is given by the relation
\begin{displaymath}
 P\simeq \frac{\beta_{g}}{4}
\end{displaymath}
while in the  weak coupling ($\beta_{g} \gg 1$) we have
\begin{displaymath}
 P\simeq 1 - \frac{3}{D\beta_{g}}
\end{displaymath}
where $\beta_{g}$ denotes the (unique) gauge coupling.\\
The five dimensional isotropic pure SU(2) model is known to undergo a first order phase transition at approximately $\beta_{g}\simeq 1.63$ from a strong coupling confining phase $\bold{S}$, where the plaquette
behaves as $\frac{\beta_{g}}{4}$, to a Coulomb phase $\bold{C_5}$ where the plaquettes asymptotically behave as $1-\frac{3}{5\beta_{g}}$,
bold line $C$ in the phase diagram \cite{Creutz,KAKF}.  The phase diagram of the
anisotropic SU(2) model, with couplings ($\beta$, $\beta^{'}$), consists of the two aforementioned phases, strong confining phase $\bold{S}$ (for small values of $\beta$, $\beta^{'}$) and Coulomb 5D phase $\bold{C_{5}}$
(for large values of $\beta$, $\beta^{'}$),
plus a new phase $\bold{W}$ for large $\beta$ and small values of $\beta^{'}$
\footnote{The so called "Layer" phase in the anisotropic 5d pure U(1) case \cite{DFV}.}.
The phase $\bold{W}$ is the continuation of the weak coupling phase
of the four dimensional  SU(2), for non zero $\beta^{'}$. In this phase the transverse-like plaquette
$P_{5}$ behaves like $\frac{\beta^{'}}{4}$ and the space-like plaquette $P_{S}$ takes the values
$1-\frac{3}{4\beta}$ indicating a four dimensional space time.\\
In the rest
of this paper we will try to determine the nature of the transitions between the three possible phases of the model.

\subsection{Order of the Phase Transitions}

We begin with the passage from the strong coupling confining phase $\bold{S}$ to the ``Layer'' one \footnote{For the case of the ``Layer'' phase we use the term loosely.}, denoted by $\bold{W}$, horizontal shadow thick line in the phase diagram.
From the perspective of the transverse plaquette the transition goes unnoticed since
both phases exhibit confinement along the
transverse direction. Thus it comes as no surprise that $P_{5}$ remains fixed at the strong coupling limit value $\frac{\beta^{'}}{4}$ throughout
the transition. As described above the mean value of the space plaquette $P_{S}$ for $\beta^{'}\neq0$ has the same value with the plaquette for the 4d SU(2) in
the same coupling $\beta$. The
space plaquette retains its value for a wide range of $\beta^{'}$ with no volume dependence.
In Fig.~2 we show the $\beta$-dependence of the susceptibility for the space plaquette
at the point $\bold{A}$ in Fig.~1, corresponding to $\beta^{'}=0.60$, for three different volumes (L=6,8 and 10). The peak of the space plaquette susceptibility $S(\hat{P}_{S})$
has the same value for the various volumes, so this transition is not really a phase transition. This line corresponds to a crossover from the
strong coupling phase $\bold{S}$ to the weak coupling phase $\bold{W}$. Both regions describe the same physics with confinement
at large distances and asymptotic freedom at small distances. The $\bold{W}$
region is a continuation of $\beta^{'}=0$ behavior to $\beta^{'}\neq0$.\\

The next step involves the study of the Strong-Coulomb transition, $\bold{S}$ to $\bold{C_{5}}$, represented by the bold line $\bold{C}$ in the phase diagram of Fig.~1.
Information on the transition is contained in both the space and transverse plaquettes. We study the transition along lines of constant $\beta$ and also
along constant $\beta^{'}$.\\
We present in Fig.~3(a) the histograms for the distribution $N(\hat{P}_{5})$ of the transverse plaquette $\hat{P}_{5}$ at the value $\beta=2.10$ which is located just below the Strong-Weak coupling phase boundary.
As it is immediately apparent we are faced with a first order phase transition. In fact from the moment that we go from L=4 to L=6 the system,
for the pseudo-critical value $\beta_{c}^{'}$, spends its time in one or the other phase with no passages between the two. The same results are obtained for bigger values of
$\beta^{'}$ on the bold phase line $\bold{C}$ in Fig.~1, indicating a strong first order bulk phase transition which separates the phases $\bold{S}$, $\bold{C_{5}}$. As we move to bigger
values of $\beta^{'}$ on the phase line $\bold{C}$ the required volume to see signal for a first order bulk phase transition is increased \cite{KnechtliRago, Ejiri}. Note that the
transition weakens as we move in the phase diagram to the left for smaller values of $\beta^{'}$, bigger values of $\beta$.\\
In Fig.~3(b) we present the histogram for the transverse plaquette for
the point $\bold{T}$, $\beta=2.50$, in the phase diagram. The lattice volume used is $16^5$ and $\beta^{'}=0.8695$ on the maximum of transverse plaquette susceptibility,
see Fig.~4(a). The two states in the histogram of Fig.~3(b) are barely distinguishable. As the transition weakens the two states come closer.
This in turn results in a smaller barrier required by the system to overcome, in order to go from one state to the other, and consequently it does not spend a substantial amount
of time in  either one of them but rather goes back and forth between them for a small volume. Furthermore, the range of a distribution is proportional to the inverse of the square root of the volume
of the system under study. For the particular volume the gap is of the same order of magnitude as the distributions range resulting in overlapping histograms for the two distinct phases.\\

\begin{figure}[!ht]
\begin{center}
\includegraphics[scale=0.45,angle=360]{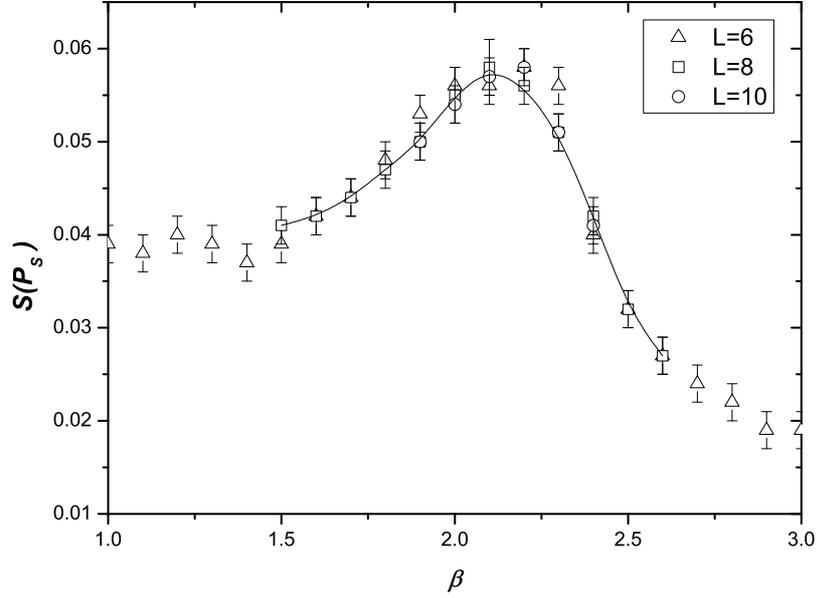}
\caption{The susceptibility $S(\hat{P}_{S})$ of the space plaquette ($\hat{P}_{S}$) at $\beta^{'}=0.60$ for three different values of lattice length (L=6, 8 and 10) as a function of $\beta$.
The curve has been drawn to guide the eye.}
\label{scbgt060}
\end{center}
\end{figure}
\begin{figure}[!h]
\begin{center}
\subfigure[]{\includegraphics[scale=0.28,angle=360]{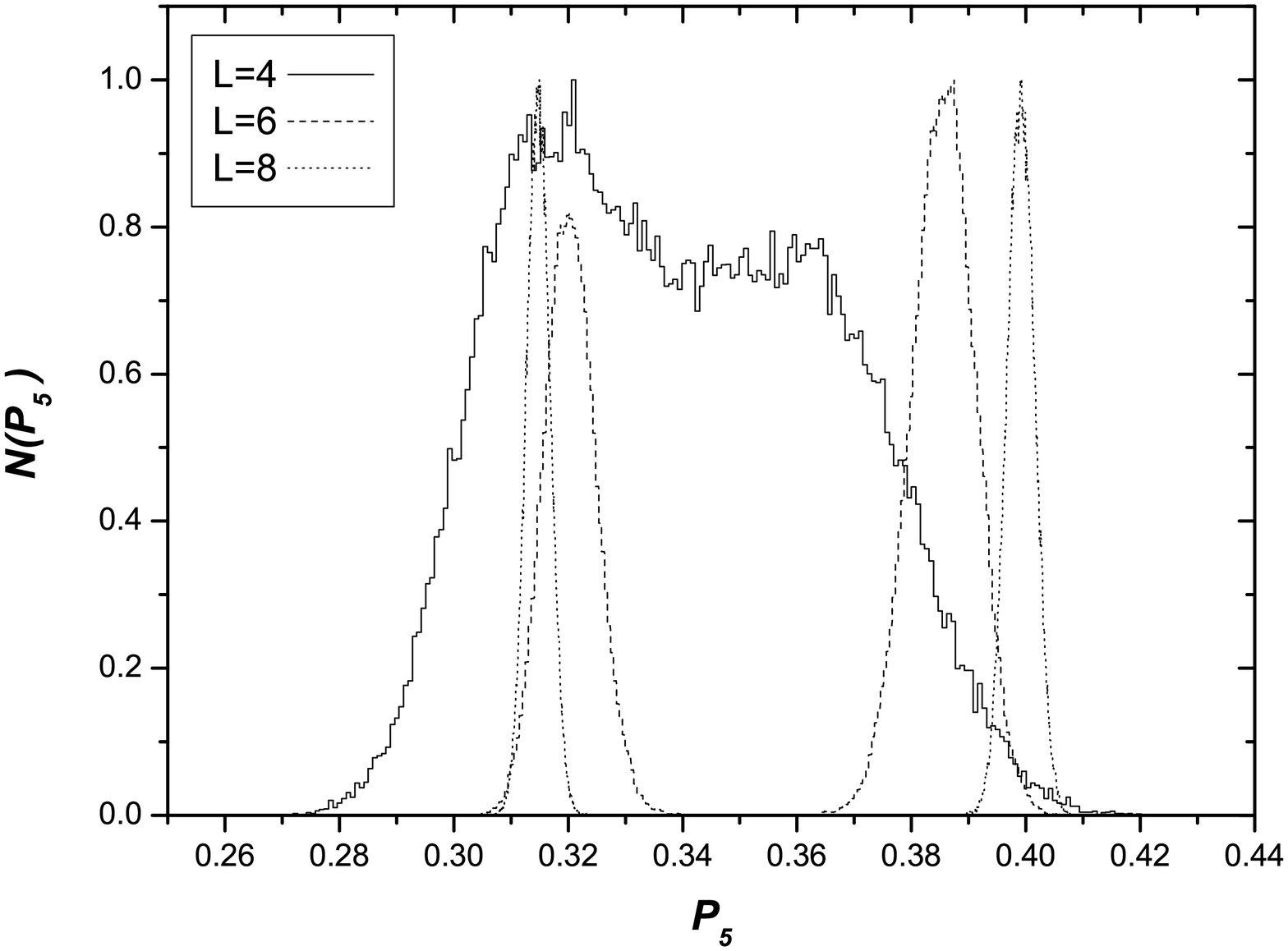}}
\subfigure[]{\includegraphics[scale=0.28,angle=360]{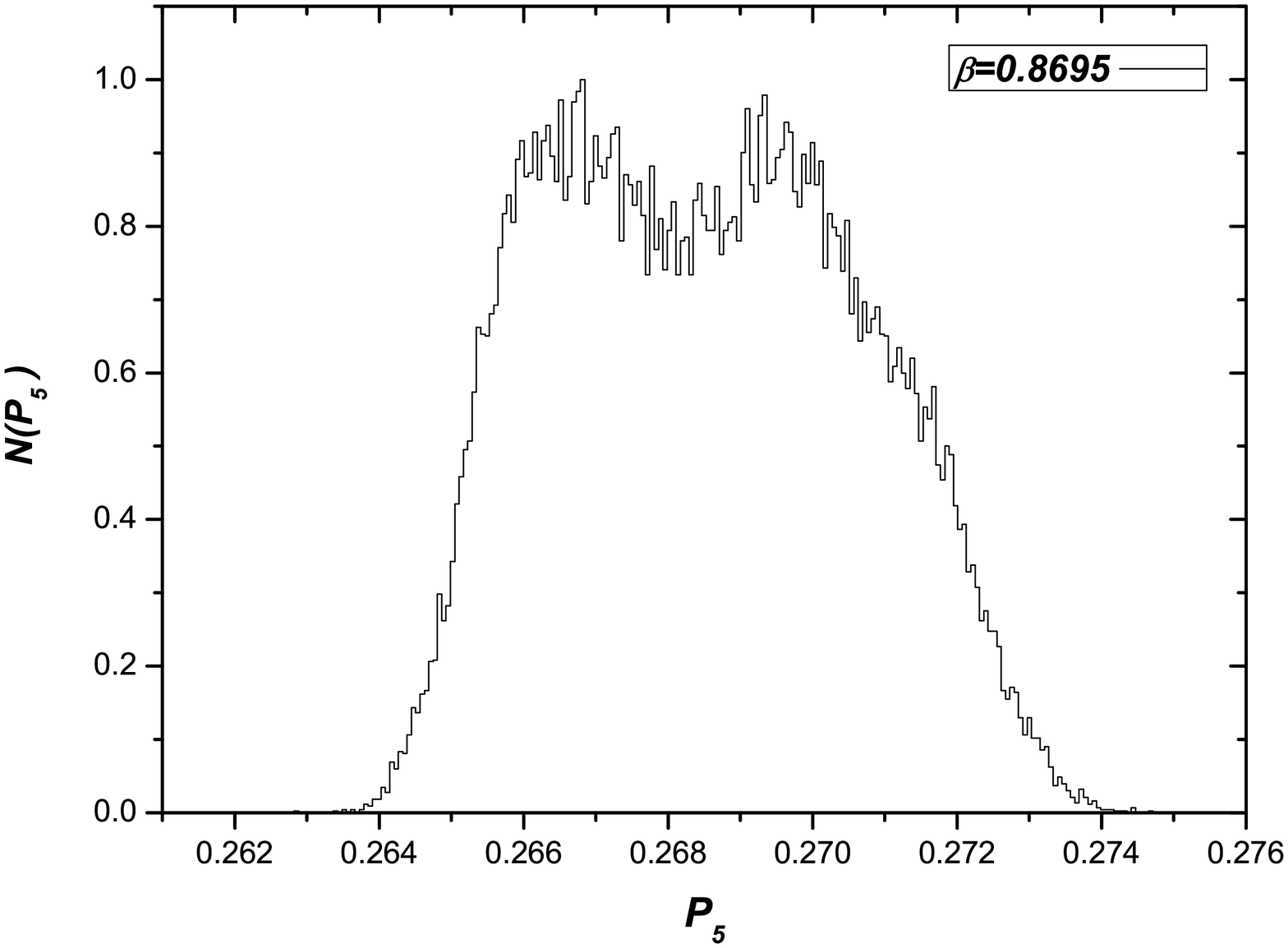}}
\caption{Panel (a) contains the histograms for the distribution $N(\hat{P}_{5})$ of the transverse plaquette $\hat{P}_{5}$ for $\beta=2.10$ and $\beta^{'}=1.070, 1.093, 1.098$
for L=4, 6 and 8 correspondingly. Panel (b) contains the histogram for $\beta=2.50$ and $\beta^{'}=0.8695$ (on the maximum of the transverse plaquette susceptibility). The lattice length is L=16.}
\label{histbgs}
\end{center}
\end{figure}

For the point $\bold{T}$, $\beta=2.50$, we present also in Fig.~4 the transverse plaquette susceptibility $S(\hat{P}_{5})$ for various lattices $(L=10,12,14 \textnormal{ and } 16)$ as a function of $\beta^{'}$ Fig.~4(a), and the maximum,
$S_{max}$, of the transverse plaquette susceptibility versus $L$ in Fig.~4(b). For the smaller volumes under study (up to L=14) the system appears to undergo a continuous phase transition. The values of the
plaquette change smoothly with $\beta^{'}$ and the maximum of the susceptibility grows with a much smaller rate than one would expect from a first order transition. It is only at L=16
that the first indications of a first order transition appear. A faint two peak signal in the distribution of the transverse plaquette (Fig.~3(b)) accompanied by an abrupt change in the values
of the susceptibility (Fig.~4). For a first order phase transition one expects the susceptibility to scale with the lattice volume. We observe indeed that the volume ratio is
$(\frac{16}{14})^{5}\simeq2$ and the obtained value for the transverse susceptibility gives
$\frac{S_{max}(L=16)}{S_{max}(L=14)}\simeq3$. We expect that this tendency for increasing the susceptibility in proportion to the lattice volume will persist for even bigger lattices.
Compared to the smaller volumes under study the susceptibility for L=16 resembles more a delta function clear signal of a first order phase transition.\\

\begin{figure}[!h]
\begin{center}
\subfigure[]{\includegraphics[scale=0.28,angle=360]{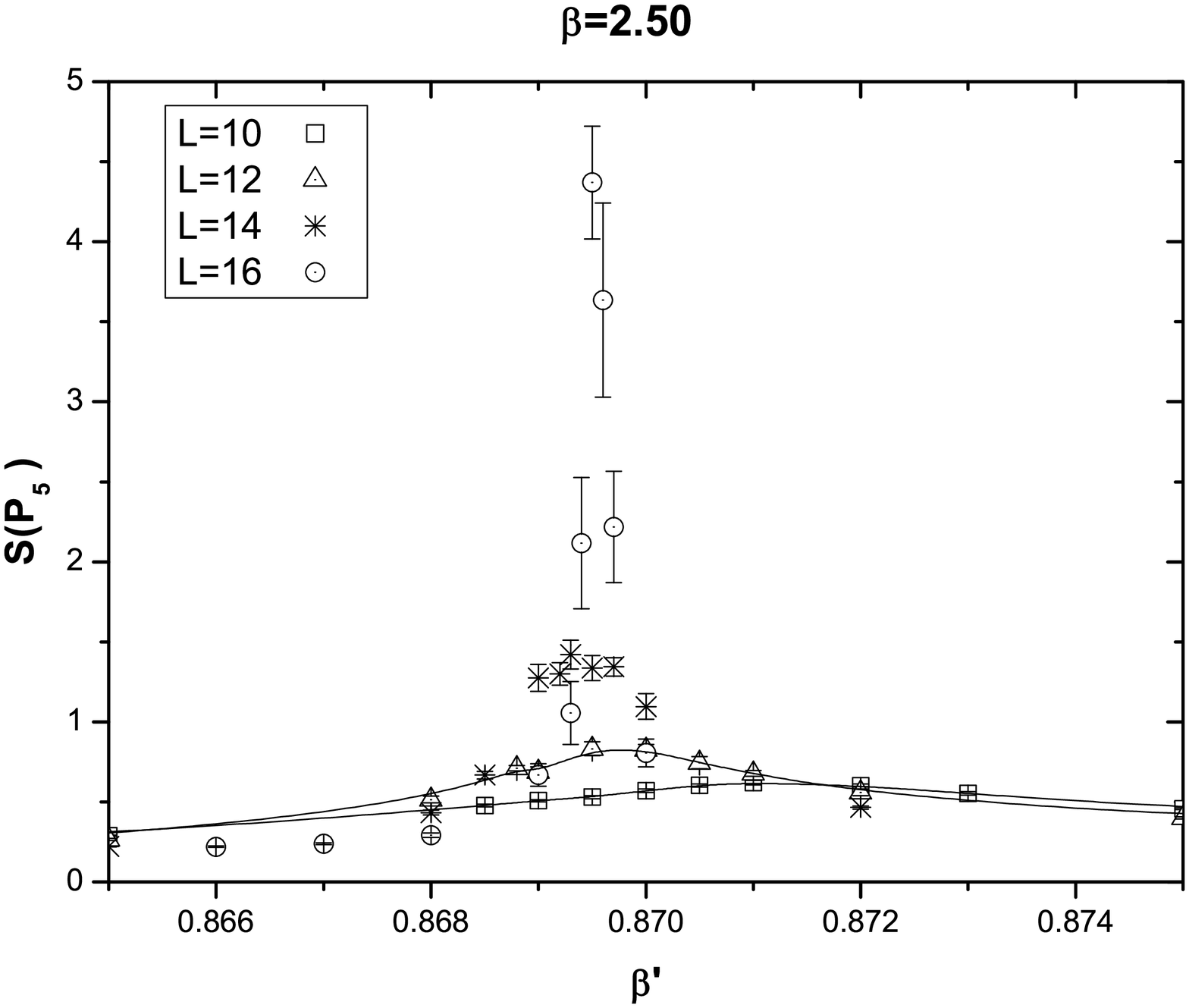}}
\subfigure[]{\includegraphics[scale=0.28,angle=360]{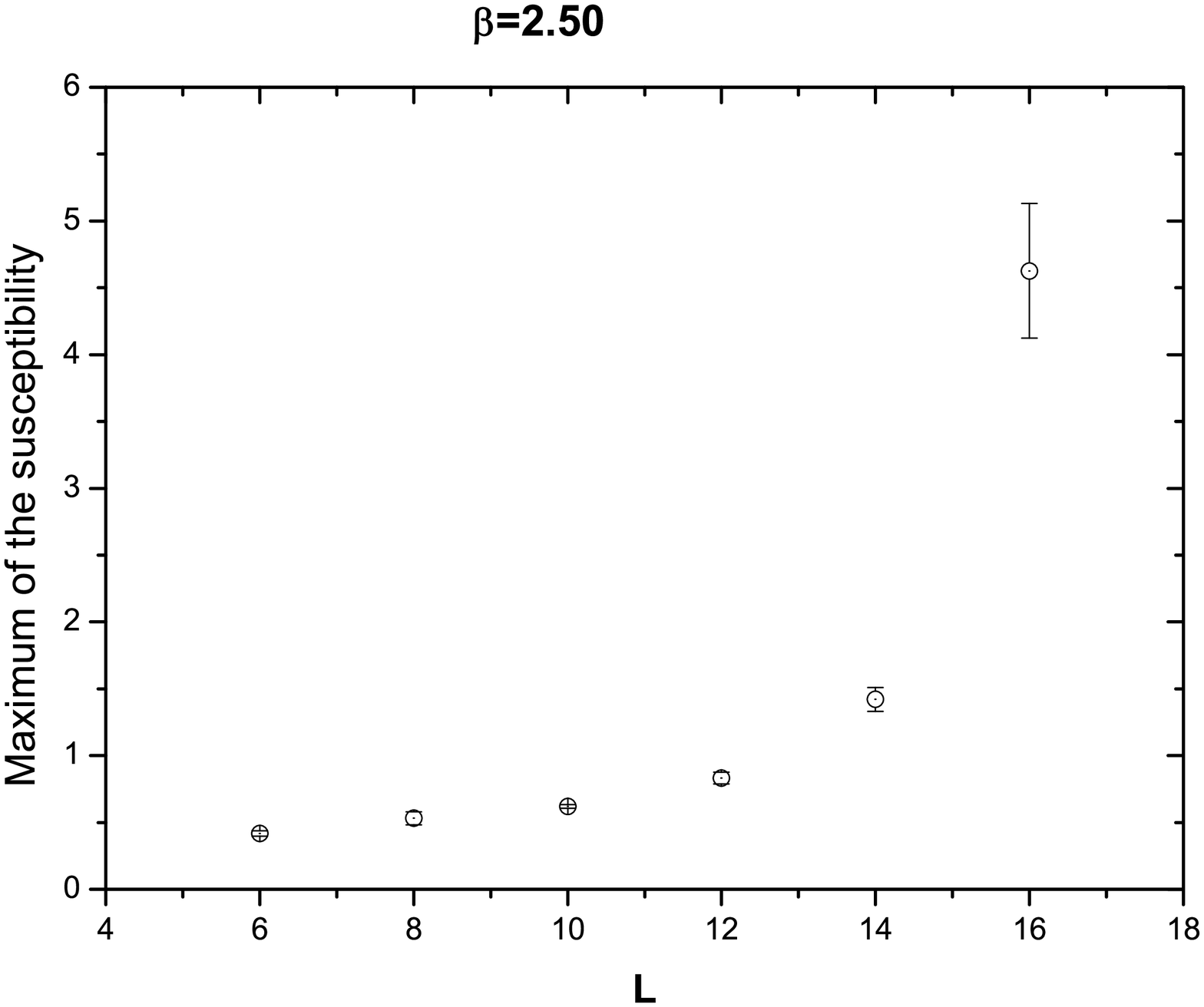}}
\caption{(a) Susceptibility $S(\hat{P}_{5})$ for the transverse plaquette at $\beta=2.50$ as a function of $\beta^{'}$ for various lattices, (b) maximum $(S_{max})$ of the transverse plaquette susceptibility $S(\hat{P}_{5})$ versus lattice length, L. The curves in panel (a) have been drawn to guide the eye.}
\label{histbgs}
\end{center}
\end{figure}

Define the transverse plaquette gap $G_{5}$ as the difference for the mean values for the transverse plaquette in the Coulomb phase $\bold{C_{5}}$ and the confining phase $\bold{S}$, $G_{5}=P_{5}(\bold{C_{5}})-P_5(\bold{S})$. We calculate the gap $G_{5}$ on the first order phase line $\bold{C}$ in the phase diagram (Fig.~1).
The gap $G_{5}$ is presented in Fig.~5 versus $\beta$ for various lattice volumes.
Each point in Fig.~5 represents a couple $(\beta^{'}, \beta)$ on this first order line.
Various volumes were used in order to acquire a signal for a first order
transition. For the cases where a clear two state signal is present in the distribution of the transverse plaquette the gap was calculated using independent Gaussian fits involving a close range of values around the two peaks. When a two peak signal could not be found we used extensive
runs originating from the two different phases, with different initial values for the fields. As a result, for the same values of the parameters ($\beta^{'}$,$\beta$), the
system was found in one or the other phase depending on the initial conditions, providing us with an estimate for the gap. The behavior of the plaquette gap versus
$\beta$ is quite interesting. It appears to increase in value until a plateau is reached in the region close to $\beta\simeq\beta^{'}$. Afterwards, and as $\beta$ is increased further, its
value is decreased rather rapidly, possibly reaching zero for values of $\beta\simeq 2.60$.\\

\begin{figure}[ht!]
\begin{center}
\includegraphics[scale=0.45,angle=360]{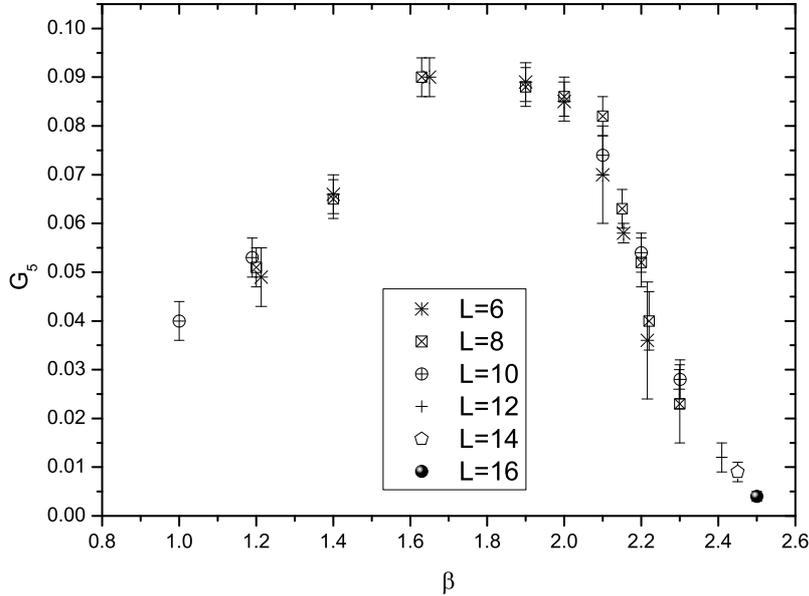}
\end{center}
\caption{Transverse plaquette gap $G_{5}$ versus $\beta$ along the first order phase line $\bold{C}$
in the phase diagram, for various lattice volumes. The last point on the right for $\beta=2.50$ corresponds to the point $\bold{T}$
in the phase diagram.}
\label{}
\end{figure}

Of particular interest is the transition between the weak coupling phase $\bold{W}$ and the Coulomb (or deconfined) phase $\bold{C_5}$. It has been suggested in \cite{Irges} that
there is a line of second order phase transitions on the phase boundary. Our findings support
the claim for a phase transition of second order or higher. We study in detail up to lattice volume $16^{5}$ two values of $\beta$. First point $\bold{B_{1}}$ with $\beta=3.00$ deep inside the weak coupling
region of SU(2) and second, point $\bold{B_{2}}$ with $\beta=2.60$ near the expected, from our results, end of the first order phase transition.\\

For the first point $\bold{B_{1}}$ of the phase diagram, at $\beta=3.00$ the relevant observables
are the transverse plaquette $P_{5}$ which should experience the passage from a confining behavior to a Coulomb one, and the transverse plaquette
susceptibility $S(\hat{P}_{5})$. For our study we used
six volumes (L=6, 8, 10, 12, 14 and 16) with the number of measurements ranging from $10^{5}$ to $2\times 10^{5}$ especially near the critical region.
Our results for the transverse
susceptibility $S(\hat{P}_{5})$ are summarized in
Fig.~6. The plaquette $P_{5}$ itself has a weak volume dependence in the critical region and
the plaquette distributions at the critical points lack a two peak signal, so a first order phase transition is excluded at least up to this lattice volume.\\
In the sequel we investigate if our results are compatible with a second order transition.
For a second order phase transition we expect that near the critical
point the correlation
length is given by the relation
\begin{equation}
 \xi\sim |\beta^{'}-\beta_{c}^{'}|^{-\nu}
\end{equation}
We assume that the pseudo-critical value of the transverse gauge coupling scales with the lattice length according
to the expression \cite{DFV}:
\begin{equation}
\beta_{c}^{'}(L)=\beta_\infty^{'}(1+C_{1}L^{-\frac{1}{\nu}})
\end{equation}
or equivalently:
\begin{equation}
\ln|\beta_{c}^{'}(L)-\beta_\infty^{'}|=C_{2}-\frac{1}{\nu}\ln(L)
\end{equation}
The asymptotic scaling law for the susceptibility peak takes the form
\begin{equation}
 S_{\textnormal{max}}(L)=C_{3}+C_{4}L^{\frac{\gamma}{\nu}}
\end{equation}

 The susceptibility $S(\hat{P}_{5})$ as a function of the five dimensional volume is depicted in Fig.~6. The susceptibility peaks display a weak
scaling with the volume. The pseudocritical $\beta^{'}_{c}$ and the maxima of the susceptibility, $S_{\textnormal{max}}(\hat{P}_{5})$, for each
lattice volume have been estimated using the multihistogram method for all the points around the critical value and are given in Table 1.

\begin{table}
\begin{center}
\begin{tabular}{ccc}
\hline \hline
 $L$ & $S_{max}(\hat{P}_{5})$ & $\beta^{'}_c$  \\
\hline
$6$  & 0.418(06) & 0.8034(15) \\
$8$  & 0.461(06) & 0.7942(08) \\
$10$ & 0.497(07) & 0.7895(04) \\
$12$ & 0.524(08) & 0.7868(02) \\
$14$ & 0.568(12) & 0.7856(02) \\
$16$ & 0.574(03) & 0.7840(02) \\
\hline
\end{tabular} \vspace*{0.3cm}
\caption{\label{beta_prime_01} The pseudocritical values of $\beta_{c}^{'}$ and the corresponding values for the peak of the transverse plaquette susceptibility $S(\hat{P}_{5})$ for $\beta=3.00$.}
\end{center}
\end{table}

Using Eq.~(7) and Table 1 we obtain (Fig.~7(a)) the following values :
\begin{displaymath}
\nu=0.57(4) \quad \textnormal{and}\quad \beta^{'}_{\infty}=0.779(1)
\end{displaymath}
for the infinite volume value of $\beta^{'}$ and the critical exponent $\nu$ with $\chi^{2}=1.27$.

An estimate for the other critical exponent $\gamma$ is given by :
\begin{displaymath}
 \gamma=0.27(2) \textnormal{ with }\chi^{2}=0.42
\end{displaymath}

In Ref.\cite{Irges,FdF}, see also \cite{Ejiri}, the authors conclude that the second order phase transition may belong to the trivial 4d Ising universality class. Under this assumption
we modify the relevant equations to include a logarithmic adjustment:
\begin{equation}
 \beta_{c}^{'}(L)=\beta^{'}_{\infty}+C_{0}(L(\ln(L))^{\frac{1}{12}})^{-\frac{1}{\nu}}
\end{equation}
The above equation, which is valid for the 4D Ising model (Ref.\cite{BJM,LM}), incorporates logarithmic corrections as opposed to the ``naive'' power-law
behavior of Eq.~(7). From Eq.~(10) and Table 1 we have:
 \begin{displaymath}
\nu=0.60(9) \quad \textnormal{and}\quad \beta^{'}_{\infty}=0.779(1)
\end{displaymath}
with $\chi^{2}=1.27$, no significant change in the quality of the fit is observed and the results are comparable with the previous values for $\nu$ and $\beta^{'}$.\\
The scaling of the susceptibility is also modified accordingly:
\begin{equation}
 S_{\textnormal{max}}(L)\approx(L(\ln(L))^{\frac{1}{4}})^{\frac{\gamma}{\nu}}
\end{equation}
Using the previous value of $\nu$ and Table 1 we have:
\begin{displaymath}
 \gamma=0.25(17) \textnormal{ with } \chi^{2}=0.45.
\end{displaymath}

The values that we obtain are very far from those of the 4D Ising model
\footnote{In the 4d Ising model $\nu=0.50$ and $\frac{\gamma}{\nu}=2.00$.}.
Both methods give a rough estimate $\frac{\gamma}{\nu}\sim 0.45$, indicating that the susceptibility scales with the
lattice volume like a square root.
If we do fix the value of $\frac{\gamma}{\nu}$ to the 4d Ising model value and repeat the fit for the maxima of
the transverse susceptibility
$S_{\textnormal{max}}(\hat{P}_{5})$, the result is bad with a big chi-square.
We estimate that we need more statistics and bigger lattice volumes to be able to have a definite
answer for the universality class of this phase transition.\\

\begin{figure}[!h]
\begin{center}
\includegraphics[scale=0.45,angle=360]{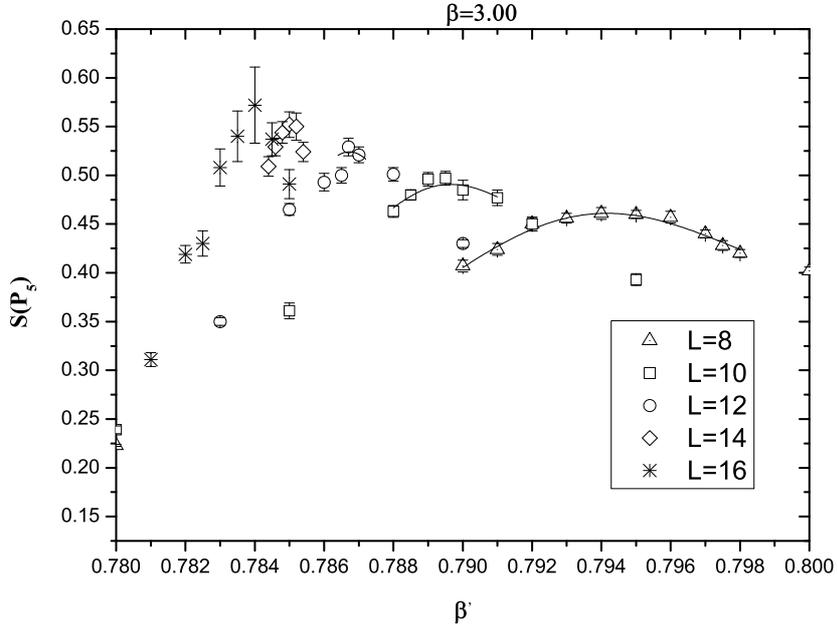}
\caption{The volume dependence for the transverse plaquette susceptibility $S(\hat{P}_{5})$ for $\beta=3.00$ as a function of $\beta^{'}$. The curves have been produced using the
multihistogram method. We observe a weak scaling with the lattice volume.}
\label{scTbg300}
\end{center}
\end{figure}
\begin{figure}[!h]
\begin{center}
\subfigure[]{\includegraphics[scale=0.28,angle=360]{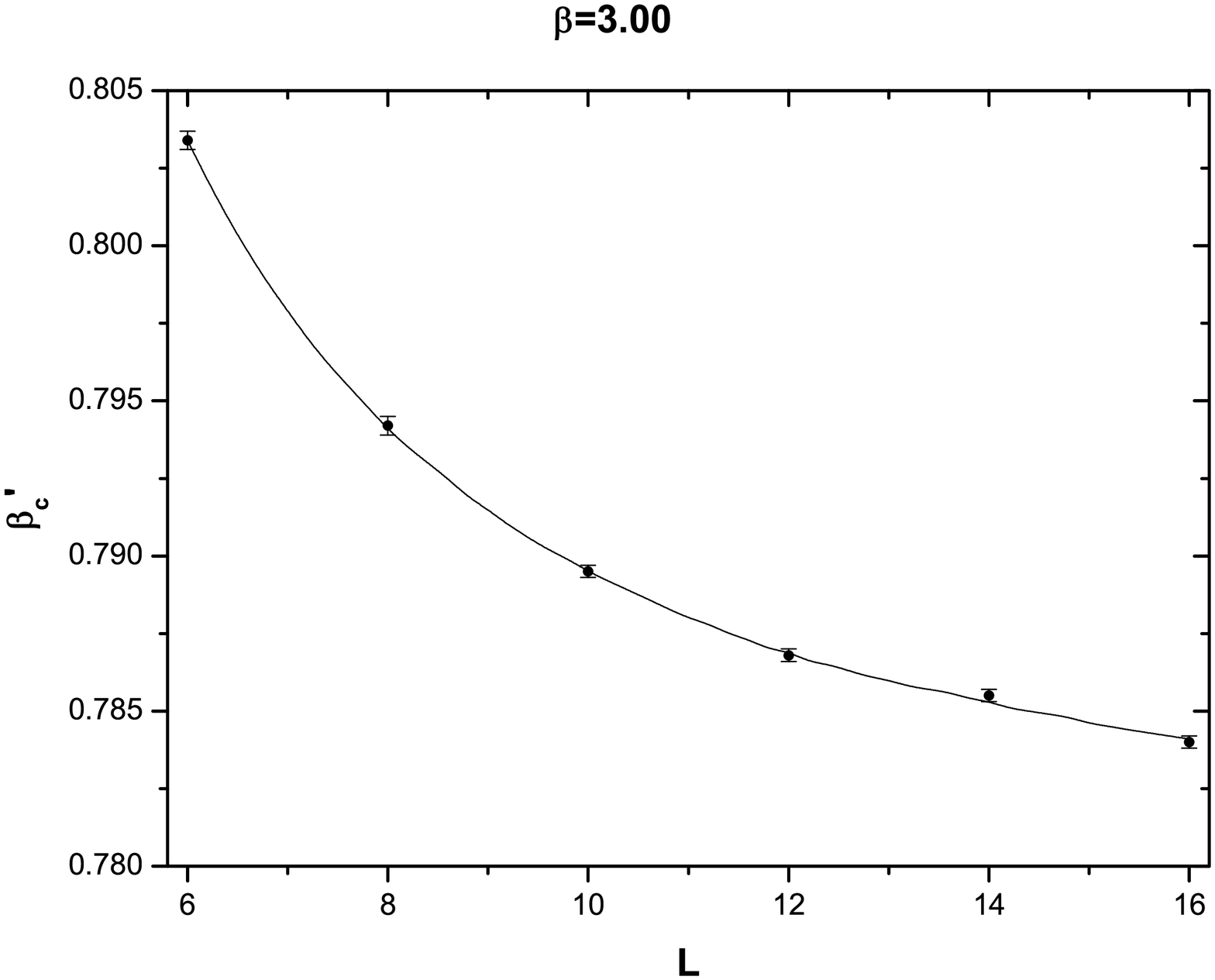}}
\subfigure[]{\includegraphics[scale=0.28,angle=360]{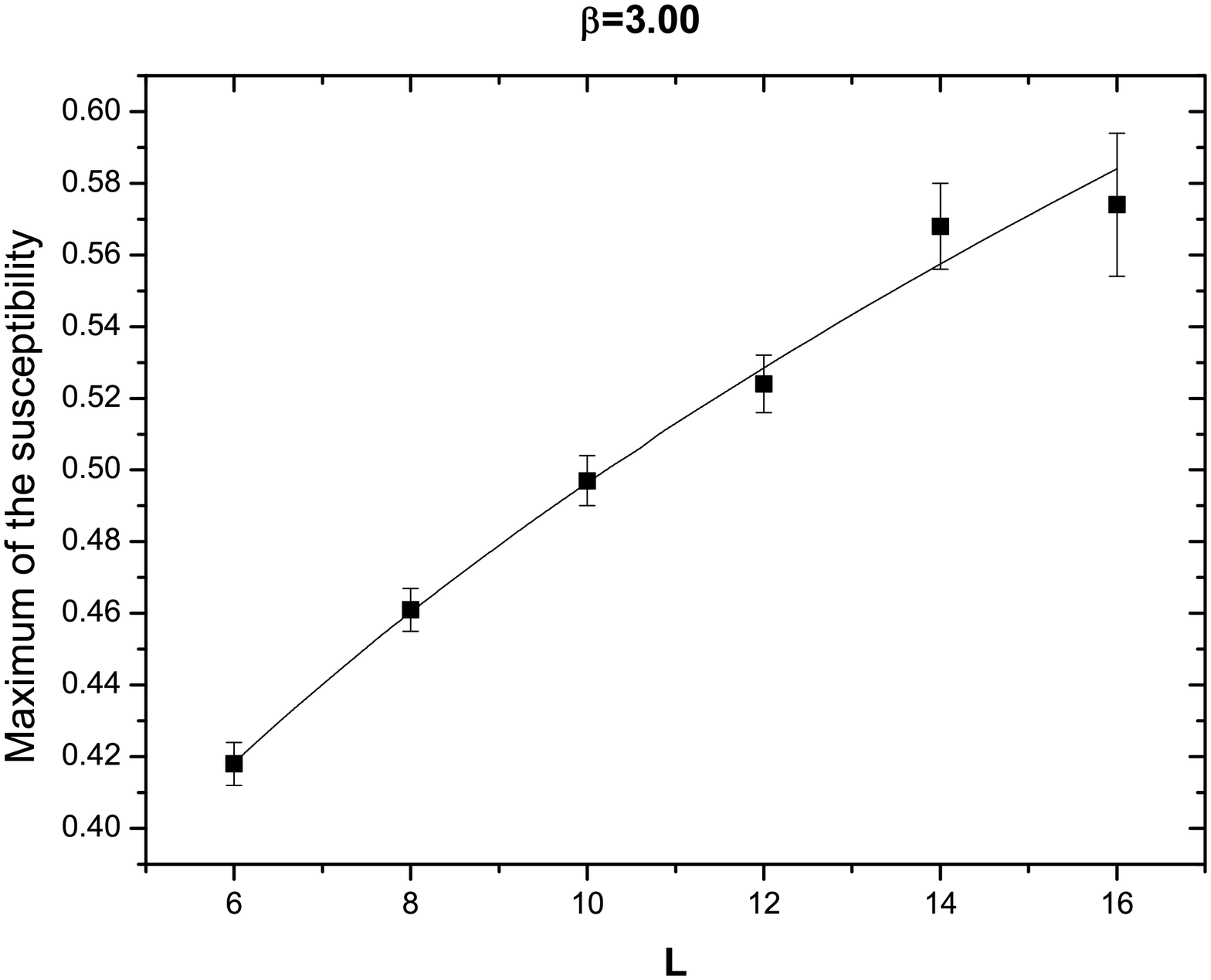}}
\caption{The pseudocritical values $\beta^{'}_{c}$, ((a)), and the maximum of the transverse plaquette susceptibility $S(\hat{P}_{5})$, ((b)), as a function of the lattice length L for $\beta=3.00$. The two different fitting
lines in panel (a), based on Eq.~(7) and Eq.~(10), are indistinguishable for this range of the lattice length. In the right hand panel (b) the same applies for the two fitting lines of Eq.~(9) and
Eq.~(11).}
\label{BcSmax}
\end{center}
\end{figure}

Next we set the value of the space coupling to $\beta=2.60$ while we allow $\beta^{'}$ to vary, point $\bold{B_{2}}$ in the phase diagram. We repeat the previous analysis for the new value
of $\beta$ and the results are presented in Table 2.

\begin{table}
\begin{center}
\begin{tabular}{ccc}
\hline \hline
 $L$ & $S_{max}(\hat{P}_{5})$ & $\beta^{'}_c$  \\
\hline
$6$  & 0.394(18) & 0.8580(10) \\
$8$  & 0.465(08) & 0.8525(05) \\
$10$ & 0.531(09) & 0.8480(05) \\
$12$ & 0.597(09) & 0.8455(04) \\
$14$ & 0.712(30) & 0.8445(02) \\
$16$ & 0.827(40) & 0.8440(02) \\
\hline
\end{tabular} \vspace*{0.3cm}
\caption{\label{beta_prime_01} The pseudocritical values of $\beta_{c}^{'}$ and the corresponding values for the peak of the transverse plaquette susceptibility $S(\hat{P}_{5})$ for $\beta=2.60$}
\end{center}
\end{table}

From  Table 2 and equations (7) and (9) we have:\\
\begin{displaymath}
 \nu=0.33(4) \qquad \gamma=0.84(14) \qquad \beta^{'}_{\infty}=0.843(1)
\end{displaymath}
with $\chi^{2}=0.25$ and  $\chi^{2}=0.51$ respectively and the ratio $\frac{\gamma}{\nu}=2.54(53)$. \\
If we use Eqs.~(10) and (11) we have for the critical exponents and the infinite volume limit of $\beta^{'}$:
\begin{displaymath}
\nu=0.34(4) \qquad \gamma=0.77(13) \qquad \beta^{'}_{\infty}=0.843(1)
\end{displaymath}
with $\chi^{2}=0.25$ and $\chi^{2}=0.52$ respectively and the ratio $\frac{\gamma}{\nu}=2.26(47)$. The above results are obtained after the exclusion of the point L=6
from the analysis. The ratio $\frac{\gamma}{\nu}$, from the two methods, is comparable within errors. We observe from our results that up to L=16 the scaling of the susceptibility gives
no  indication for the presence of a first order phase transition.
\begin{figure}[!h]
\begin{center}
\includegraphics[scale=0.50,angle=360]{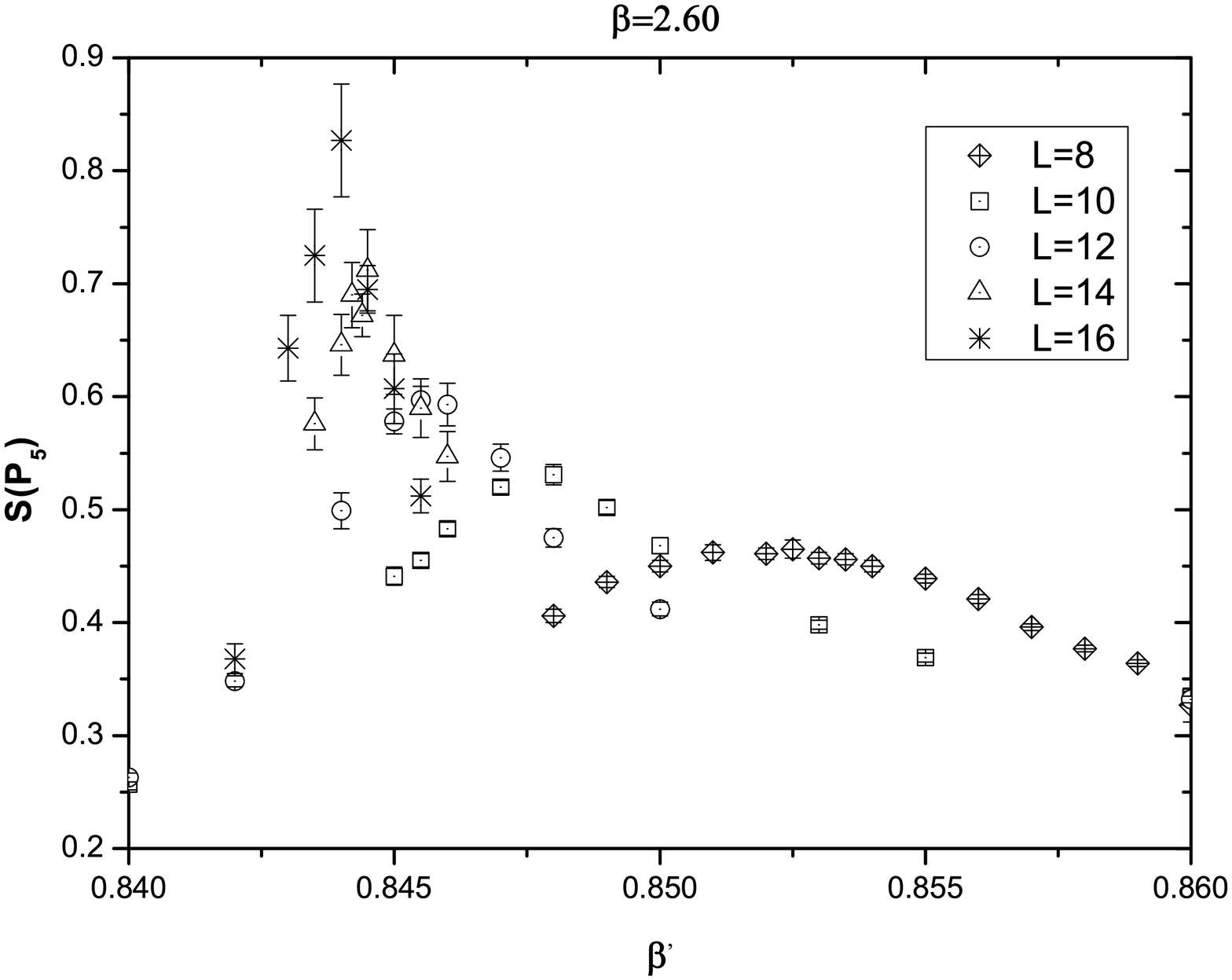}
\caption{The volume dependence for the transverse plaquette susceptibility $S(\hat{P}_{5})$, for $\beta=2.60$ as a function of $\beta^{'}$.
We used various volumes from L=8 to L=16.}
\label{scTbg300}
\end{center}
\end{figure}

\begin{figure}[!h]
\begin{center}
\subfigure[]{\includegraphics[scale=0.28,angle=360]{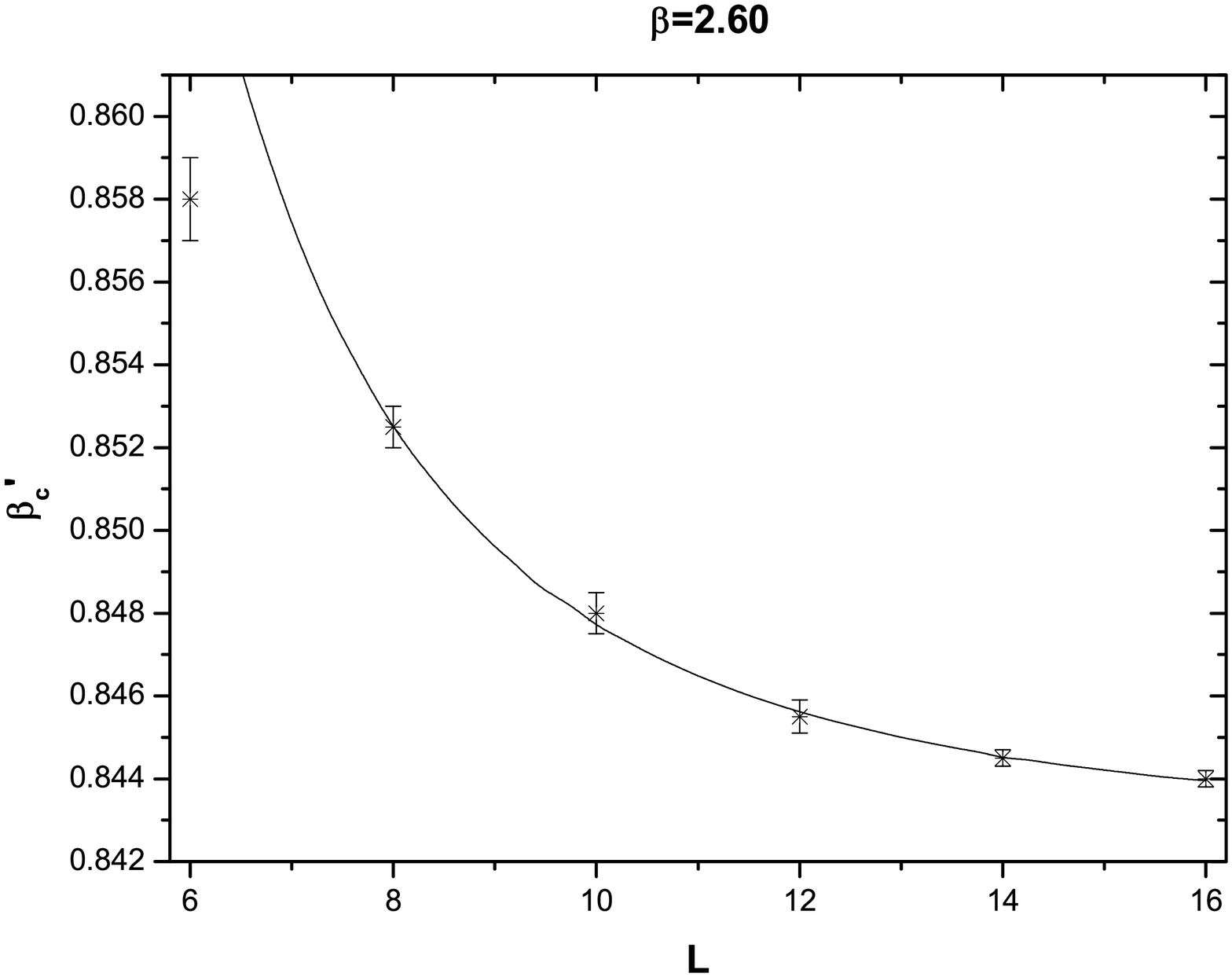}}
\subfigure[]{\includegraphics[scale=0.28,angle=360]{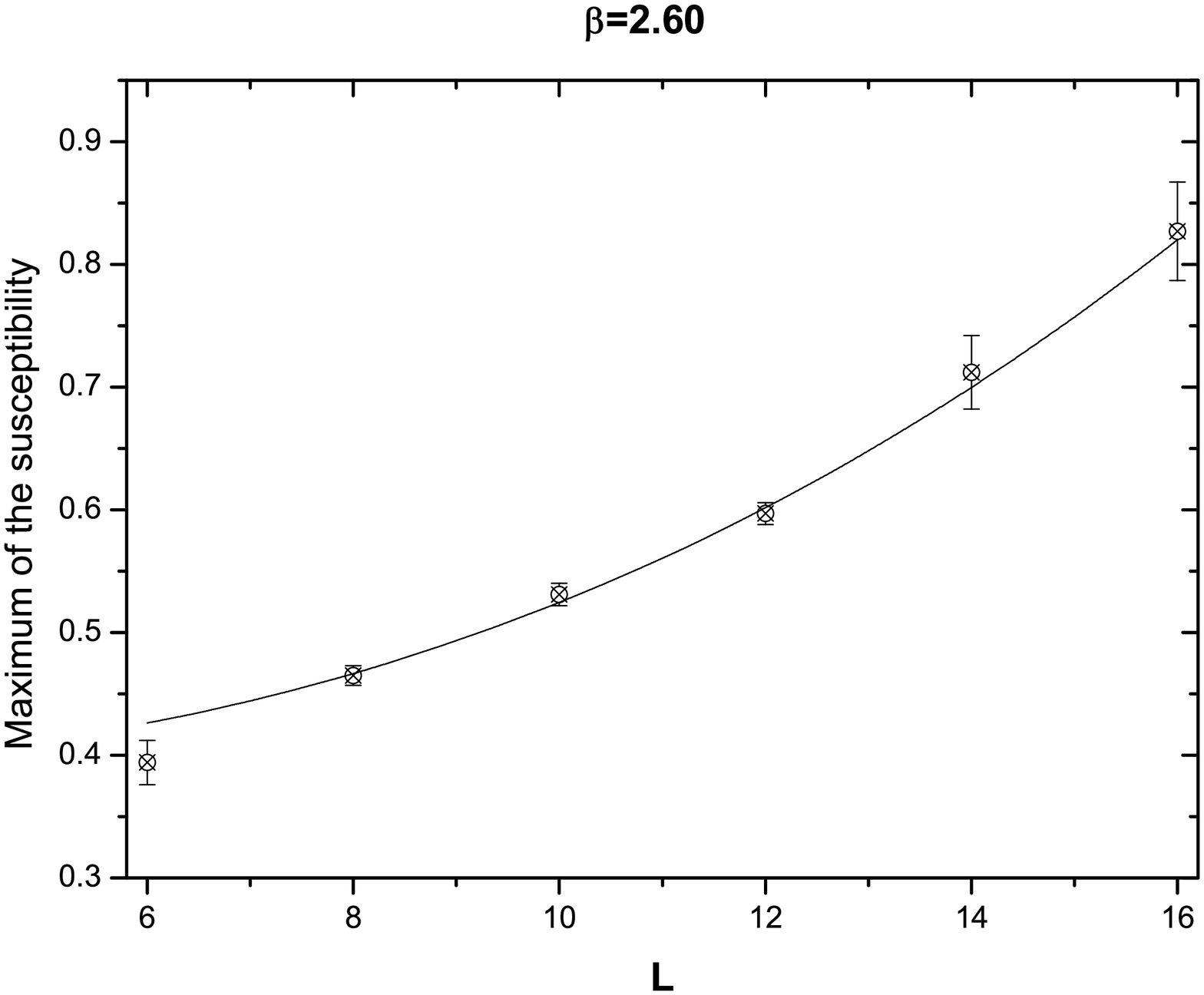}}
\caption{The pseudocritical values $\beta^{'}_{c}$, ((a)), and the maximum of the transverse plaquette susceptibility $S(\hat{P}_{5})$, ((b)), as a function of the lattice length L for $\beta=2.60$.
The two different fitting lines in (a) and (b), based on Eq.~(7), Eq.~(9) and Eq.~(10), Eq.~(11), are indistinguishable for this range of the lattice length.}
\label{BcSmax}
\end{center}
\end{figure}

\newpage

\section{Discussion}
We consider a pure SU(2) gauge model in 4+1 dimensions with anisotropic couplings. The model appears to have
two dinstict phases: The 5d Coulomb phase $\bold{C_5}$ for large $\beta$, $\beta^{'}$ and the continuation of the 4d SU(2) to $\beta^{'}\neq0$, $\bold{S}$ and $\bold{W}$ regions in the $\beta$, $\beta^{'}$ plane.\\

The study of the phase transitions shows that the strong coupling $\bold{S}$ phase of SU(2) for $\beta^{'}>0$ and $\beta<2.20$ is
separated from the weak coupling $\bold{W}$ phase by a crossover.\\

\begin{table}[!h]
\begin{center}
\begin{tabular}{|c|c|c|c|c|c|}
\hline
$\beta$ & Fit & $\nu$ & $\gamma$ & $\gamma/\nu$ &
$\beta_{\infty}^{\prime}$\tabularnewline
\hline
\hline
2.60 & Eqs.(7) \& (9) & 0.33(4) & 0.84(14) & 2.54(53) & 0.843(1) \tabularnewline
\cline{2-6}
  & Eqs.(10) \& (11) & 0.34(4) & 0.77(13) & 2.26(47) & 0.843(1) \tabularnewline
\hline
\hline
3.00 & Eqs.(7) \& (9) & 0.57(4) & 0.27(02) & 0.47(06) & 0.779(1) \tabularnewline
\cline{2-6}
  & Eqs.(10) \& (11) & 0.60(9) & 0.25(17) & 0.41(24) & 0.779(1) \tabularnewline
\hline
\end{tabular}
\caption{ Results for the critical exponents for  $\beta=2.60$ and
$\beta=3.00$ using fit functions given by Eqs. (7) and (9),  and Eqs. (10)
and (11).}
\label{Cr_exp_results}
\end{center}
\end{table}

The phase transition between the $\bold{S}$ and $\bold{C_5}$ phases
is a strong first order phase transition, that becomes weaker as the coupling $\beta$ increases and
approaches the value $\beta=2.50$, point $\bold{T}$ in the phase diagram.
To corroborate this result we study the plaquette gap across this first order line and present the transverse plaquette gap
$G_{5}$ in Fig.~5, where we can see clearly that $G_{5}$ decreases as $\beta$ approaches
the point $\bold{T}$ in the phase diagram, probably going to zero in the region $2.50 \lesssim\beta\lesssim 2.60$.\\

Furthermore for the phase transition from the weak coupling $\bold{W}$ phase to 5d Coulomb $\bold{C_5}$ we provide detailed evidence for a continuous phase transition of second order type or higher. We concentrate our attention in two points
$\bold{B_{1}}$, $\bold{B_{2}}$ for $\beta=3.00$ and $\beta=2.60$, deep in the weak coupling phase of
4D SU(2) and near the vanishing of the plaquette gap correspondingly. In this study lattice volume varies from $L=6$ to $L=16$. We summarize our work in Fig.~10, where we compare the behavior of the transverse
plaquette susceptibility for the points $\bold{B_{1}}$ and $\bold{B_{2}}$ versus $L$. It is clear that we have
a different scaling with the lattice size. The curve for $\beta=2.60$ is increasing roughly like the square of $L$ with
positive convexity while for $\beta=3.00$ is increasing like the square root of $L$ with negative convexity.
In Table 3 we summarize our results for the critical exponents $\nu, \gamma$, $\frac{\gamma}{\nu}$ and $\beta^{'}_{\infty}$ for $\beta=2.60$ and $\beta=3.00$.
It is possible that the point $\bold{B_{2}}$
lies on the border region where the order of phase transition changes.
Moreover there is always the possibility that the transition at $\bold{B_{1}}$ turns to a crossover, but we cannot conclude on that scenario.
Anyway up to this volume we have no indication for a linear dependence by the lattice volume $L^{5}$ so a first
order phase transition is excluded for both cases.
If this conclusion persists for even bigger volumes it would give rise to a non trivial field theory in the continuum limit. It was noted in Ref.\cite{Irges,FdF} that near the transition, the system reduces dimensionally with the low energy degrees of freedom those of a four dimensional Georgi-Glashow type model.\\

\begin{figure}[!h]
\begin{center}
\includegraphics[scale=0.50,angle=360]{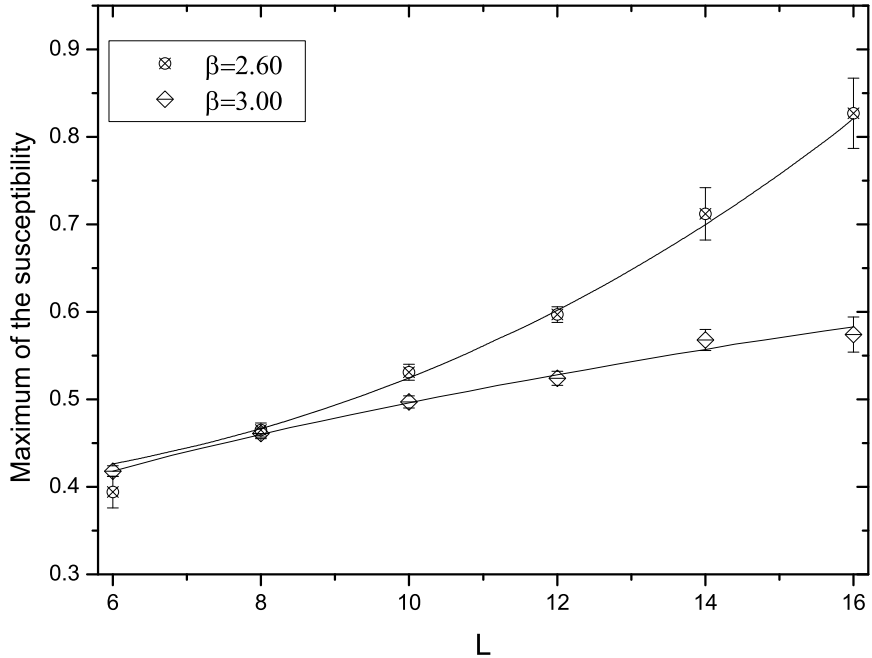}
\caption{Maximum of the transverse plaquette susceptibility $S_{\textnormal{max}}(\hat{P}_{5})$ for $\beta=2.60$ and $\beta=3.00$ versus lattice length $L$, a combination of figures 7(b) and 9(b). For $\beta=2.60$ the fitting line increases roughly like the square of $L$ while for $\beta=3.00$ it behaves as the square root of $L$ with negative convexity.}
\label{scTbg300}
\end{center}
\end{figure}

We would like to make a final comment regarding the second order phase transition. Denoting the
parameter $\widehat{\gamma}^{2}=\frac{\beta^{'}}{\beta}$, the second order phase transition in
Ref.\cite{Irges} is realized for $\widehat{\gamma}^{2}<1$ as in our case,
while in Ref.\cite{FdF, Ejiri} it seems to be a finite temperature phase transition for
$\widehat{\gamma}^{2}>1$.

\section{Acknoweledgements:} K.F. thanks N.Irges for useful discussions on the subject during the Milos conference. We like also to thank K.Anagnostopoulos,
P.Dimopoulos, G.Koutsoumbas and A.Tsapalis for reading the manuscript.


\end{document}